\documentclass[useAMS,usenatbib]{mn2e}

\usepackage{graphicx}
\usepackage{natbib}
\usepackage{lscape}
\usepackage{amssymb}

\newcommand{\oo}{\rm [O\,II]$\lambda\lambda$3727,3729/[O\,III]$\lambda$5007}
\newcommand{\ohb}{\rm [O\,III]$\lambda$5007/H$\beta$}
\newcommand{\oha}{\rm [O\,I]$\lambda$6300/H$\alpha$}
\newcommand{\nha}{\rm [N\,II]$\lambda$6583/H$\alpha$}
\newcommand{\sha}{\rm [S\,II]$\lambda$6726,6731/H$\alpha$}

\title[A survey of Low Luminosity Compact sources]{A survey of Low Luminosity 
Compact sources and its implication for evolution of radio-loud AGNs. II.
Optical analysis}

\author []{....}

\author[Kunert-Bajraszewska \& Labiano]{M. Kunert-Bajraszewska$^{1}$\thanks
{E-mail: magda@astro.uni.torun.pl} \& A. Labiano$^{2}$\\ 
$^{1}$Toru{\'n} Centre for Astronomy, N. Copernicus University,
Gagarina 11, 87-100 Toru{\'n}, Poland\\
$^{2}$European Space Agency (ESA), European Space Astronomy Centre (ESAC),
28691 Villanueva de la Ca\~{n}ada, Madrid, Spain\\ 
}

\begin{document}

\date{Accepted 1988 December 15. Received 1988 December 14; in original form 1988 October 11}

\pagerange{\pageref{firstpage}--\pageref{lastpage}} \pubyear{2002}

\maketitle

\label{firstpage}

\begin{abstract}

This is the second in a series of papers concerning a new sample of
low luminosity compact (LLC) objects. 
Here we discuss the optical properties of the sample based
on Sloan Digital Sky Survey
(SDSS) images and spectra. We have generated different diagnostic
diagrams and classified the sources as high and low excitation galaxies
(HEG and LEG, respectively). We have studied the 
jet-host interactions, relation between radio and optical line emission 
and evolution of the radio source
within a larger sample that included also the published samples of 
compact steep spectrum (CSS), gigahertz peaked spectrum (GPS) sources 
and FR\,II and FR\,I objects.
The optical and radio properties of the LLC 
sample are in general consistent with brighter CSS and 
large-scale radio sources, although the LLC objects have lower values of
[O\,III] luminosity than the more powerful CSS sources ($L_{\rm 1.4\,GHz}>
10^{25}~{\rm W~Hz^{-1}}$). 
However, when LLC are added to the other samples,
HEG and LEG seem to follow independent, parallel evolutionary tracks. 
Regarding ionization mechanisms,
LLC and luminous CSS objects behave like FR\,II sources, while
FR\,I seem to belong to a different group of objects. 
Based on our results, we propose the independent,
parallel evolutionary tracks for HEG and LEG sources, evolving from GPS -
CSS - FR.
\end{abstract}


\section{Introduction}

The compact radio sources consist of two population of objects:
the gigahertz-peaked spectrum (GPS) and compact steep spectrum (CSS)
sources. These are considered to be young and evolve
into large radio objects, FR\,I/FR\,II 
%\citep{fr74} during their lifetimes 
\citep[][for a review]{f95, r96, odea98}.
The GPS sources are considered to be entirely contained within the
extent of the narrow-line region ($\leq$ 1~kpc). 
Unbeamed, symmetric GPS sources have been classified as Compact
Symmetric Objects by \citet{wilkinson}.
CSS sources are thought
to extend to the size of the host galaxy ($\leq$ 20~kpc).
Compact radio sources are the ideal for learning more
about the relation between formation and evolution of the
host galaxy, the trigger of the activity and its effect on the
nuclear regions and ISM of the host galaxy.

Once the nuclear activity and radio source are triggered, the small-scale
jets expand through the natal cocoon, driving outflows in the emission line
gas (fast outflows and jet-cloud interactions). In some cases the
interaction of the radio jets with the ISM can disrupt the jet and change 
the morphology and luminosity of the source \citep{kb07}. 
In large radio sources, the emission line activity is connected with black hole mass,
fuelling mechanism and type of the accreating gas
\citep{Hardcastle07, butti10}.
However the morphological division of large radio objects into FR\,I and
FR\,II does not correspond to low/high excitation division: FR\,Is show
typically faint optical nuclei and low excitation spectra, while among
FR\,IIs we have both low and high excitation galaxies. The spectroscopic analysis of GPS/CSS sources
- progenitors of large FR\,I/FR\,II objects - can allow to derive their
accretion properties at times close to the jet launching. 

In GPS/CSS sources, the jet is still crossing the ISM and the interaction 
with the ISM is stronger than in large radio sources.
Observations of the ionized gas in GPS and CSS sources show the presence of
such interactions \citep{holt09, holt, labiano}.
Therefore, other ionization mechanisms must be taken into account, such as 
jet-induced shocks or precursor photoionization.
Furthermore, some radio sources show traces of star formation associated to
the AGN event \citep{chiaberge, allen, labiano08} and young
star photoionization should also be taken into account.

It has been showed that emission line luminosities are correlated with radio
power of large scale sources \cite{raw89,butti10} and small CSS and GPS
objects \citep{morganti97, labianolet}. Moreover  
\citet{labianolet} has found that the GPS and CSS
sources (galaxies and quasars) show a strong correlation between
[O\,III] line luminosity and size of the radio source. 
However, this correlation may not be present in fainter radio sources where 
the nuclear ionization source can dominate over the effects of the weaker 
jets on the ISM.  
There are indications \citep{tasse}, based on the environmental studies of
radio sources
at moderate redshifts (z$\sim$~0.5-1), that low-luminosity radio AGNs  lie in
denser environments than the powerful objects, have lower stellar masses and
show excess mid-IR emission consistent with a
hidden radiatively efficient active nucleus. \citet{tasse} also argue that
the low luminosity radio activity is associated with the re-fuelling of
massive black holes.
Most of the GPS and CSS sources known so far are
powerful objects and the clue to solve the evolution puzzle can be the analysis of less
powerful compact sources.

In this paper we will investigate the characteristics of the optical properties 
of 29 out of 44 low luminosity CSS sources selected from the FIRST survey.
We compare
these with more powerful compact objects and FR\,I and FR\,II large sources.

\section{Optical data}
\label{obs}
The sample of Low Luminosity Compact (LLC) sources has been selected from
the FIRST survey and observed with MERLIN at L-band and C-band. The selection criteria 
and the radio properties of the sample were discussed and analysed in \citet{kunert10}, 
hereafter Paper\,I.  
All of the LLC sources are nearby objects with redshift $z<0.9$.

Optical data are available for 29 LLC sources and have been obtained from the Sloan Digital 
Sky Survey (SDSS) Data Release 7 \citep[DR7, ][]{Abazajian}. DR7 includes fits 
for all the emission lines in the spectra. However, the SDSS pipeline has problems 
fitting multi-component lines such as H$\alpha$+[NII] doublet or lines with both broad 
and narrow components. In these cases, we fitted the lines using TOPCAT and the Splot task in 
IRAF. 

DR7 data are not corrected of Galactic extinction. We have applied the \cite{cardelli} 
extinction law to correct for it, using the E(B-V) values listed in NED. 
Table~\ref{lines} lists the line fluxes, corrected for Galactic extinction, 
for each source with available SDSS spectra.

To improve our statistics, we completed the sample of compact sources by
taking CSS/GPS sources with available spectroscopic and radio data from the
literature: 28 CSS/GPS sources were
taken from \citet{labiano07} and 8 low luminosity CSSs were taken from
\citet{buti09}. Moreover we performed spectroscopic analysis od CSS objects
from \citet{f2001} and \citet{mar03} samples ({\it combined sample},
Table~\ref{combined}) and discussed them in Appendix A. Their radio properties were
already used in the statistical studies presented in Paper\,I. However, 
the SDSS data are available for only 14 objects from the {\it combined
sample}.
To explore evolution with 
size (i.e. age) we also included the [O\,III] line luminosities 
of available 114 large sources from the revised 3C sample of FR\,I and FR\,II sources
\citep[LRL sample:][and Willott et al.
1999\footnote{http://www-astro.physics.ox.ac.uk/~sr/grimes.html
}]{laing83, buti}. 

Throughout the paper, we assume a cosmology with
${\rm H_0}$=71${\rm\,km\,s^{-1}\,Mpc^{-1}}$, $\Omega_{M}$=0.27,
$\Omega_{\Lambda}$=0.73 \citep{Spergel03}. Distances were calculated using the astronomy 
calculator by \cite{wright}.

\section{Notes on individual sources}
\label{notes}
In this section we describe the main features in the SDSS optical images and
spectra of the LLC sample.
The HEG/LEG classification is based on the line ratios observed in the
SDSS spectra and the \cite{butti10} definitions. Our sample shows that all
sources with
log \oo $\la 0.2$ are LEG while those with log \oo $\ga 0.2$ are  HEG
(Figure \ref{BPT} and
Table \ref{lines}). The data suggests that the \ohb \, can be used to
distinguish between HEG and LEG:
HEG show log \ohb $\ga 0.75$ while LEG show  log \ohb \, $\ga 0.75$. Based
on these criteria, we assigned a
HEG/LEG classification for those sources where
we could measure only the \oo \, ratio. These sources are marked with `*'.

\noindent {\bf 0025+006}.
HEG. The source shows extended ($\sim$20\,kpc) emission and a bright
nucleus. The image
suggest a possible tail towards NW.

\noindent {\bf 0754+401}.
HEG. The image shows an extended ($\sim$10\,kpc) source with a bright
nucleus with a possible
arm or tail at $\sim$10\,kpc towards the NE.

\noindent {\bf 0810+077}.
LEG. Very elongated ($\sim$40\,kpc, NW-SE) galaxy with a bright nucleus,
possibly a
disk galaxy. The radio source is perpendicular to the longest optical axis
(Paper\,I). This 
behaviour has been observed before in e.g. 3c 236 \citep{odea01,tremblay10}.

\noindent {\bf 0821+321}.
Very extended source ($\sim$60\,kpc) with a faint nucleus. The image shows
two bright
emission regions towards NE (at $\sim$15\,kpc) and SW (at $\sim$25\,kpc),
embedded in the
extended emission.

\noindent {\bf 0835+373}.
Compact source with a faint nucleus.

\noindent {\bf 0846+017}.
HEG. Diffuse source with a faint nucleus. The image suggest possible
extended emission towards W
($\sim$20\,kpc).

\noindent {\bf 0850+024}.
The image shows extended emission ($\sim$30\,kpc) suggesting an elongated
(SE-NW)
galaxy or an optical jet. However, the radio jet is oriented EW (Paper\,I). 
The 
spectrum is 
consistent with AGN or stellar continuum. It seems to have a Ly$\alpha$
break but it is
on the border of the spectrum. The emission line ratios suggest possible HII
(star
forming) galaxy. However, this is based on limits.

\noindent {\bf 0851+024}.
LEG*. The image shows a diffuse and faint source with a bright compact
object at
$\sim$50\,kpc SE. 
The optical sources are not necessarily related.

\noindent {\bf 0907+049}.
The image shows two bright merging sources embedded in extended emission.
The separation between the nuclei amounts to ~20 kpc.
The extended emission has a size of $\sim$50\,kpc.
The Western
component shows an elongated NE-SW structure while the East component 
is a compact object. The probable radio jet however, is aligned NS    
(Paper\,I). The optical 
image shows a third, faint source south of the system at $\sim$60\,kpc with
some diffuse
emission between them. \cite{koylov95} already suggested that 0907+049 was a
member of a
group of galaxies.

\noindent {\bf 0914+504}.
The optical image shows a compact source. The spectra shows broad
[Mg\,II]$\lambda$2800 and a powerlaw 
continuum consistent with a QSO.

\noindent {\bf 0921+143}.
LEG. Large system ($\sim$50\,kpc) with a bright nucleus. The image shows two
bright, compact
regions at $\sim$20\,kpc towards W and E, and a fainter compact region at
$\sim$30\,kpc SW. 
The image suggests a merging system in a very crowded field.

\noindent {\bf 0923+079}.
HEG*. The image shows two close-by ($\sim$10\,kpc) bright compact sources.
The radio emission
corresponds to the E component. The spectra shows a flat continuum with
broad H$\beta$ and   
possibly H$\alpha$. However the former is right in the border of the
spectrum. The
spectrum 
therefore suggests a Sy1 or QSO. There are also hints of faint
[Mg\,II]$\lambda$2800 and [NeV]$\lambda$3425 emission lines.  
However, the S/N is too low in this region. The radio maps show also a very
compact source,
consistent with the optical classification (Paper\,I).

\noindent {\bf 0931+033}.
LEG. The optical image shows a bright $\sim$20\,kpc source with a bright
nucleus with a small
($\sim$7\,kpc) tail towards SE, and very elongated, jet-like $\sim$40\,kpc
emission towards NW.
The radio map, however, shows a compact source with no jet present
(Paper\,I).

\noindent {\bf 0942+355}.
HEG. The optical image shows a compact source ($\sim$12\,kpc) with weak
extended emission surrounding
it and a faint tail towards NW. The orientation of the tail is aligned with
the radio jet (Paper\,I). 

\noindent {\bf 1007+142}.
LEG. The image shows a very extended source ($\sim$50\,kpc) with a bright
nucleus and a bright
compact region at $\sim$8\,kpc for the center, towards W. However, the radio
structure (2.3\,kpc) is oriented NS (Paper\,I). 

\noindent {\bf 1037+302}.
LEG. Extended ($\sim$24\,kpc) asymmetric galaxy, showing a tail towards NE.
The radio
structure 
(2.6\,kpc) is however perpendicular to this structure (Paper\,I). This
behaviour has been observed
before in e.g. 3c 236 \citep{odea01,tremblay10}. The optical image also show
two compact regions on
the E and W sides of the source.

\noindent {\bf 1053+505}.
HEG*. The spectrum looks clearly as a QSO, with a strong powerlaw and bright
broad
[Mg\,II]$\lambda$2800 and H$_\beta$ emission. The image shows a compact
bight source,
consistent with the QSO classification. The radio map shows a 5.8\,kpc
structure, unresolved in the optical image (Paper\,I).

\noindent {\bf 1140+058}.
HEG*. The image shows a compact source. The spectrum shows faint powerlaw
continuum with broad
emission lines, suggesting a QSO and consistent with the core-jet radio
morphology observed   
(Paper\,I).

\noindent {\bf 1154+435}.
HEG*. The image and radio map show a compact source. The spectrum shows
clear powerlaw continuum
with bright, broad emission lines, consistent with a QSO. The emission line
ratios suggest a possible HII
(star forming) source. However, the data fall in in the HEG/HII limit.

\noindent {\bf 1156+470}.
The image shows a small source with some diffuse emission around and no
clear nucleus.
There are a few aligned knots of emission towards NW, extending
$\sim$9\,kpc. This structure
is aligned with the smaller 3.6\,kpc radio structure (Paper\,I).

\noindent {\bf 1308+451}.
Extended ($\sim$35\,kpc) source with a very complex structure and several
bright compact
regions and suggests an interacting system in a very crowded field. The
image shows a   
bright $\sim$10\,kpc NW-SE jet-like structure aligned with the smaller radio
structure (Paper\,I). 

\noindent {\bf 1321+045}.
The image shows an extended ($\sim$40\,kpc) source with a faint nucleus and
two bright knots
towards NW embedded in the extended region. The radio jet however, is
perpendicular to this 
structure (Paper\,I). This behaviour has been observed before in e.g. 3c 236
\citep{odea01,tremblay10}.

\noindent {\bf 1359+525}.
The image shows an extended galaxy ($\sim$30\,kpc) with a faint nucleus,
several bright knots
around it and a possible small ($\sim$10\,kpc) companion.

\noindent {\bf 1402+415}.
HEG*. The image shows a faint compact galaxy ($\sim$20\,kpc). The spectrum
shows no traces of AGN.

\noindent {\bf 1407+363}.
HEG. The image shows a very elongated system ($\sim$70\,kpc) consisting of
three bright regions
aligned NS and connected by diffuse, extended emission. The radio source
position corresponds  
to the central region (Paper\,I). The image also suggests a faint tail
connected to a fourth knot, SW
of the system.

\noindent {\bf 1411+553}.
The image shows a faint compact galaxy ($\sim$15\,kpc). The spectrum shows
no traces of an AGN.

\noindent {\bf 1418+053}.
The image shows a faint $\sim$15\,kpc source with two faint knots of
emission towards SW at
$\sim$22 and 30\,kpc. The spectrum shows no traces of an AGN.

\noindent {\bf 1506+345}.
The image shows an extremely extended, elongated ($\sim$45\,kpc, NE-SW)
galaxy with
resolved 
dust lanes, bright emission knots and a large ($\sim$25\,kpc) diffuse tail
towards N, leaving
the galaxy from the E, where the brightest structure is. The spectrum
however, corresponds to 
the central, fainter region. The radio map shows also a complex structure
but smaller source.
Previous observations have identified this source as an interacting 
system (Paper\,I). 

\noindent {\bf 1532+303}.
The image shows a very faint source with a possible double component. SDSS
estimates a
redshift of z=0.0009. However, the spectrum is extremely noisy and, based on
the faint small
source in the image, the redshift is probably larger.

\noindent {\bf 1542+390}.
HEG*. The image shows a very small faint source ($\sim$10\,kpc).

\noindent {\bf 1550+444}.
The image shows an extended ($\sim$20\,kpc) source with a faint nucleus and
a possible
merging companion SW. The radio map shows a 5\,kpc NS structure (Paper\,I).

\noindent {\bf 1558+536}.
LEG*. The image shows a NS elongated galaxy ($\sim$20\,kpc) at the radio
source coordinates,
with a closeby ($\sim$15\,kpc) compact, bright companion (towards NE),
suggesting interaction.
The radio map shows a 3.6\,kpc long structure, aligned with the optical
emission
(Paper\,I).

\noindent {\bf 1601+528}.
LEG*. The image shows an extended ($\sim$40\,kpc) source with bright nucleus
and a bright compact
region (at $\sim$10\,kpc towards SW) in a crowded field. 

\noindent {\bf 1610+407}.
LEG. The image shows an extended ($\sim$15\,kpc) source, slightly elongated
EW, with a bright nucleus.
The orientation is consistent with the small (2.6\,kpc) radio source
(Paper\,I).

\noindent {\bf 1641+320}.
HEG*. The image shows two bright compact sources separated by $\sim$25\,kpc,
suggesting
interaction. The spectrum shows a clear powerlaw and broad emission lines,
consistent with a QSO.
\citet{broth} identified this system as a two interacting QSO. 
The SDSS spectrum corresponds to the southern one.

Summing up, the radio-selected LLC sample shows a wide variety of optical
structures: 22 sources (63\%)
show extended emission, 13 sources (37\%) are compact or unresolved, and 16
sources (46\%) show a 
possible merger or complex structures in the image. Only 4 sources (11\%)
show radio-optical alignment.
The radio-optical alignment is usually seen in CSS
\citep[e.g.,][]{de_Vries97,de_Vries99, Axon00} and  
large radio sources \citep{Chambers87, McCarthy87}. However, the radio jets
in our sample are mostly   
small ($\la$5kpc) and too young to have affected the ISM and show 
traces of jet-induced ionization (therefore alignment).

\begin{landscape}
\begin{table}
\caption[]{Emission lines measurements and spectroscopic classification of
LLC sources}
\begin{tabular}{@{}c c c c c c c c c c c c c c c c c c c c @{}}
\hline
~~~Source & Class & {\it z} & {\it E(B-V)} &  
\multicolumn{1}{c}{[Mg II]} &
\multicolumn{1}{c}{[Ne V]} & 
\multicolumn{1}{c}{[O II]} & 
\multicolumn{1}{c}{[Ne III]} & 
\multicolumn{1}{c}{H$_\delta$} &
\multicolumn{1}{c}{H$_\gamma$} &
\multicolumn{1}{c}{[O III]} & 
\multicolumn{1}{c}{H$_\beta$} &
\multicolumn{1}{c}{[O III]} &
\multicolumn{1}{c}{[O III]} &
\multicolumn{1}{c}{[O I]} &
\multicolumn{1}{c}{[N II]} &
\multicolumn{1}{c}{H$_\alpha$} &
\multicolumn{1}{c}{[N II]} &
\multicolumn{1}{c}{[S II]} &
\multicolumn{1}{c}{[S II]}  \\
~~~Name & & & &
\multicolumn{1}{c}{$\lambda 2800$}& 
\multicolumn{1}{c}{$\lambda 3425$}& 
\multicolumn{1}{c}{$\lambda 3727$}& 
\multicolumn{1}{c}{$\lambda 3870$}& & &
\multicolumn{1}{c}{$\lambda 4363$} & &
\multicolumn{1}{c}{$\lambda 4959$} & 
\multicolumn{1}{c}{$\lambda 5007$}& 
\multicolumn{1}{c}{$\lambda 6300$}& 
\multicolumn{1}{c}{$\lambda 6548$}& & 
\multicolumn{1}{c}{$\lambda 6584$}&
\multicolumn{1}{c}{$\lambda 6716$}&
\multicolumn{1}{c}{$\lambda 6731$} \\

~~~ & & & & & &
\multicolumn{1}{c}{$\lambda 3729$}&
& & & & & & & & & & & & \\
\hline
0025+006& HEG	&	0.1	&	0.02	&	--	&	--
&	142	&	20	&	--	&	--	&	--
&	14	&	44	&	128	&	--	&	51
&	82	&	138	&	53	&	43	\\
0754+401& HEG	&	0.07	&	0.05	&	--	&	--
&	69	&	14	&	--	&	3	&	--
&	13	&	60	&	171	&	--	&	38
&	94	&	91	&	32	&	28	\\
0810+077& LEG	&	0.11	&	0.02	&	--	&	--
&	60	&	--	&	--	&	3	&	--
&	7	&	10	&	22	&	18	&	45
&	58	&	104	&	46	&	36	\\
0835+373& LEG	&	0.4	&	0.04	&	--	&	--
&	7	&	--	&	--	&	--	&	--
&	1	&	1	&	3	&	--	&	*
&	*	&	*	&	--	&	--	\\
0846+017& HEG	&	0.35	&	0.04	&	--	&	--
&	11	&	--	&	--	&	--	&	--
&	3?	&	7	&	21	&	--	&	3
&	14	&	17	&	7	&	3	\\
0850+024& --	&	0.46	&	0.04	&	--	&	--
&	14	&	--	&	--	&	--	&	--
&	4	&	4	&	12	&	--	&	1
&	15	&	2	&	2	&	2	\\
0851+024& LEG*	&	0.4	&	0.04	&	--	&	--
&	2	&	--	&	--	&	--	&	--
&	--	&	--	&	1	&	--	&	--
&	--	&	--	&	--	&	--	\\
0914+504& --	&	0.63	&	0.01	&	1	&	--
&	2	&	--	&	--	&	--	&	--
&	--	&	--	&	*	&	--	&	--
&	--	&	--	&	--	&	--	\\
0921+143& LEG	&	0.14	&	0.03	&	--	&	--
&	65	&	--	&	--	&	4	&	--
&	11	&	--	&	7	&	18	&	38
&	38	&	80	&	\multicolumn{2}{c}{81}	\\		
0923+079& HEG*	&	0.44	&	0.05	&	--	&	--
&	12	&	--	&	1	&	3	&	--
&	7n	&	18	&	53	&	--	&	--
&	--	&	--	&	--	&	--	\\
	&&		&		&		&		&
	&		&		&		&		&
74b	&		&		&		&		&
	&		&		&	\\	
0931+033& LEG	&	0.23	&	0.03	&	--	&	--
&	25	&	--	&	--	&	--	&	--
&	3	&	5	&	12	&	5	&	23
&	20	&	51	&	14	&	17	\\
0942+355& HEG	&	0.21	&	0.01	&	--	&	8
&	17	&	10	&	--	&	--	&	--
&	7	&	26	&	76	&	--	&	38
&	26	&	45	&	5	&	4	\\
1007+142& LEG	&	0.21	&	0.04	&	--	&	--
&	23	&	--	&	--	&	1	&	--
&	3	&	4	&	4	&	5	&	15
&	11	&	28	&	19	&	22	\\
1037+302& LEG	&	0.09	&	0.02	&	--	&	--
&	59	&	--	&	--	&	--	&	--
&	--	&	9	&	17	&	22	&	50
&	46	&	132	&	55	&	10	\\
1053+505& HEG*	&	0.82	&	0.02	&	114	&	--
&	5	&	--	&	--	&	--	&	18
&	4n	&	5	&	14	&	--	&	--
&	--	&	--	&	--	&	--	\\
	&&		&		&		&		&
	&		&		&		&		&
90b	&		&		&		&		&
	&		&		&	\\	
1140+058& HEG*	&	0.5	&	0.03	&	9	&	--
&	20	&	--	&	--	&	--	&	--
&	2n	&	16	&	50	&	--	&	--
&	--	&	--	&	--	&	--	\\
	&&		&		&		&		&
	&		&		&		&		&
12b	&		&		&		&		&
	&		&		&	\\	
1154+435& HII/HEG*	&	0.23	&	0.01	&	--	&	13
&	31	&	25	&	--	&	4n	&	6
&	19n	&	55	&	161	&	6	&	13
&	61n	&	30	&	\multicolumn{2}{c}{184}	\\		
	&&		&		&		&		&
	&		&		&	41b	&		&
81b	&		&		&		&		&
371b	&		&		&		\\
1308+451& --	&	0.39	&	0.02	&	--	&	--
&	12	&	--	&	--	&	--	&	--
&	--	&	--	&	--	&	--	&	--
&	--	&	--	&	--	&	--	\\
1321+045& --	&	0.26	&	0.03	&	--	&	--
&	11	&	--	&	--	&	--	&	--
&	3	&	--	&	--	&	--	&	4
&	11	&	13	&	7	&	3	\\
1359+525& --	&	0.12	&	0.01	&	--	&	--
&	14	&	--	&	--	&	--	&	--
&	--	&	--	&	--	&	--	&	11
&	31	&	72	&	\multicolumn{2}{c}{23}	\\		
1402+415& HEG*	&	0.36	&	0.02	&	--	&	--
&	3	&	--	&	--	&	--	&	--
&	--	&	2	&	4	&	--	&	--
&	19	&	--	&	--	&	--	\\
1407+363& HEG	&	0.15	&	0.01	&	--	&	--
&	16	&	--	&	--	&	--	&	--
&	--	&	9	&	25	&	7	&	14
&	25	&	29	&	\multicolumn{2}{c}{17}	\\		
1411+553& --	&	0.28	&	0.02	&	--	&	--
&	3	&	--	&	--	&	--	&	--
&	--	&	--	&	--	&	--	&	--
&	--	&	--	&	--	&	--	\\
1418+053& --	&	0.46	&	0.04	&	--	&	--
&	--	&	--	&	--	&	--	&	--
&	--	&	--	&	3	&	--	&	--
&	--	&	--	&	--	&	--	\\
1542+390& HEG*	&	0.55	&	0.02	&	--	&	--
&	3	&	--	&	--	&	--	&	--
&	--	&	3	&	7	&	--	&	--
&	--	&	--	&	--	&	--	\\
1558+536& LEG*	&	0.18	&	0.01	&	--	&	--
&	22	&	--	&	--	&	--	&	--
&	--	&	5	&	7	&	--	&	7
&	11	&	21	&	\multicolumn{2}{c}{22}	\\		
1601+528& LEG*	&	0.11	&	0.02	&	--	&	--
&	11	&	--	&	--	&	--	&	--
&	--	&	4?	&	--	&	--	&	40
&	20	&	85	&	\multicolumn{2}{c}{34}	\\		
1610+407& LEG	&	0.15	&	0.01	&	--	&	--
&	62	&	--	&	--	&	4	&	--
&	11	&	13	&	34	&	--	&	32
&	49	&	70	&	51	&	24	\\
1641+320& HEG*	&	0.59	&	0.03	&	19	&	6
&	21	&	7	&	2n	&	5n	&	3
&	12n	&	25	&	79	&	--	&	--
&	--	&	--	&	--	&	--	\\
	&&		&		&		&		&
	&		&	8b	&	15b	&		&
49b	&		&		&		&		&
	&		&		&		\\
\hline
Avg. Error& --	&	--	&	--	&	19\%	&	9\%
&	8\%	&	8\%	&	31\%	&	44\%	&	30\%
&	19\%	&	15\%	&	7\%	&	13\%	&	24\%
&	16\%	&	11\%	&	12\%	&	17\%	\\
\hline
\end{tabular}

\begin{minipage}{230mm}
Description of the columns:
(1) source name;
(2) spectroscopic classification (see text for details), '*' means the
classification is based on the \oo \, ratio only;
(3) redshift; 
(4) galactic extinction;
Rest of columns: flux for each line.
'*' - The line is present but on the edge of the spectrum (and unfittable).
'?' - Low S/N, Questionable detection.
n = narrow
b = broad
Wavelengths are in \AA. Flux in $\rm 10^{-16}~erg~s^{-1}~cm^{-2}$ ,
\end{minipage}
\label{lines}
\end{table}
\end{landscape}

\begin{figure*}
\centering
\includegraphics[width=8.5cm,height=7cm]{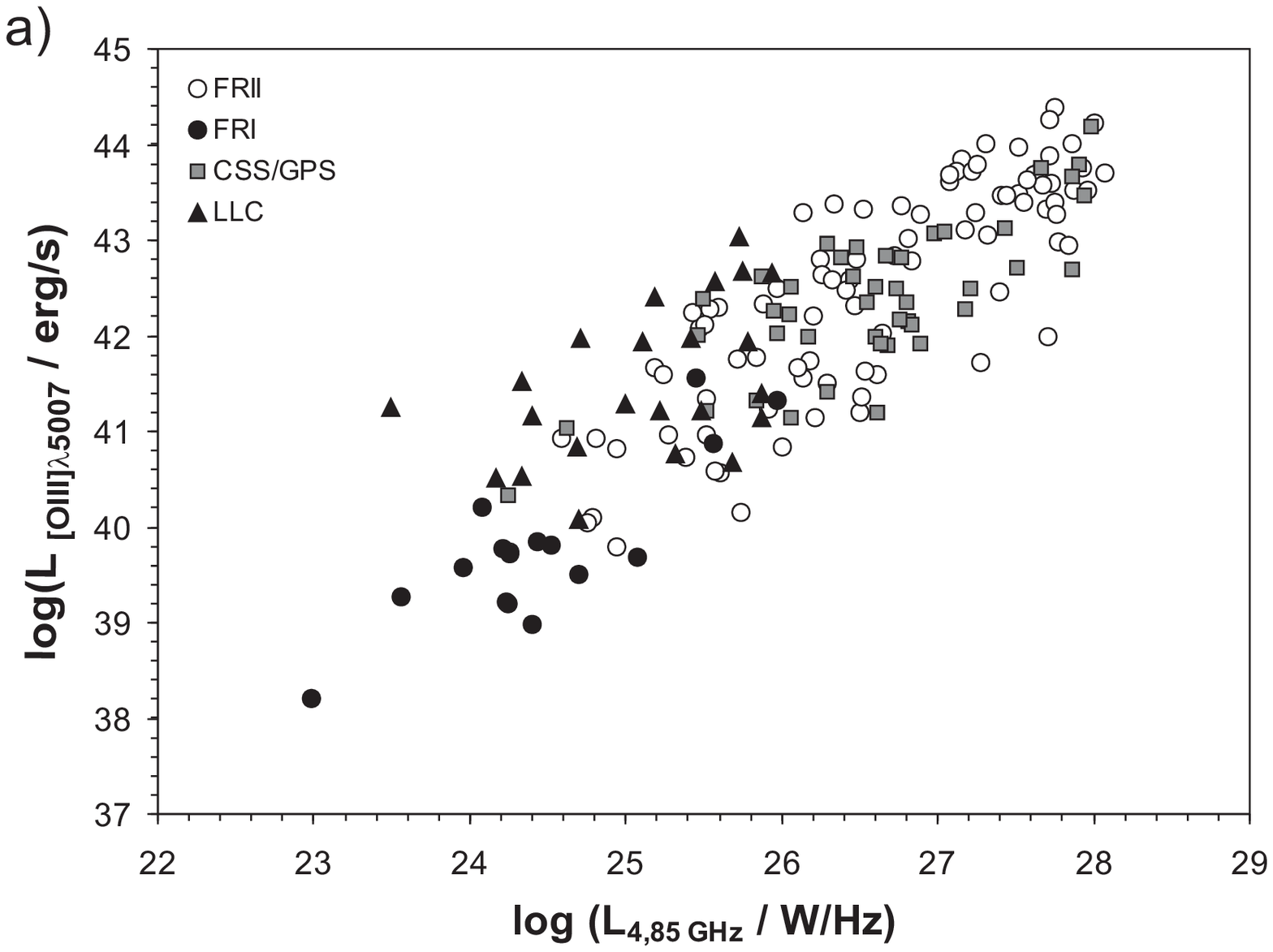}
\includegraphics[width=8.5cm,height=7cm]{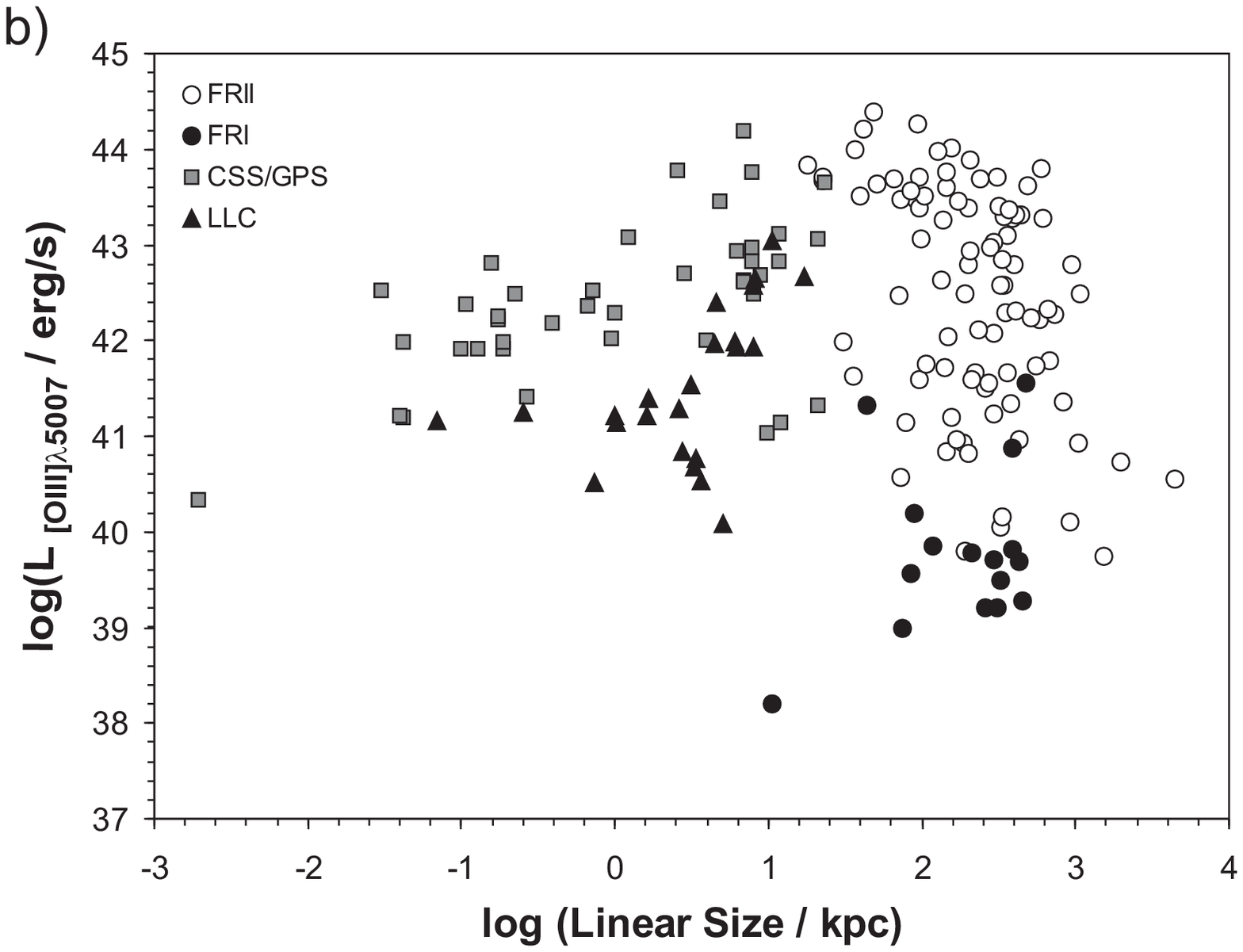}
\includegraphics[width=8.5cm,height=7cm]{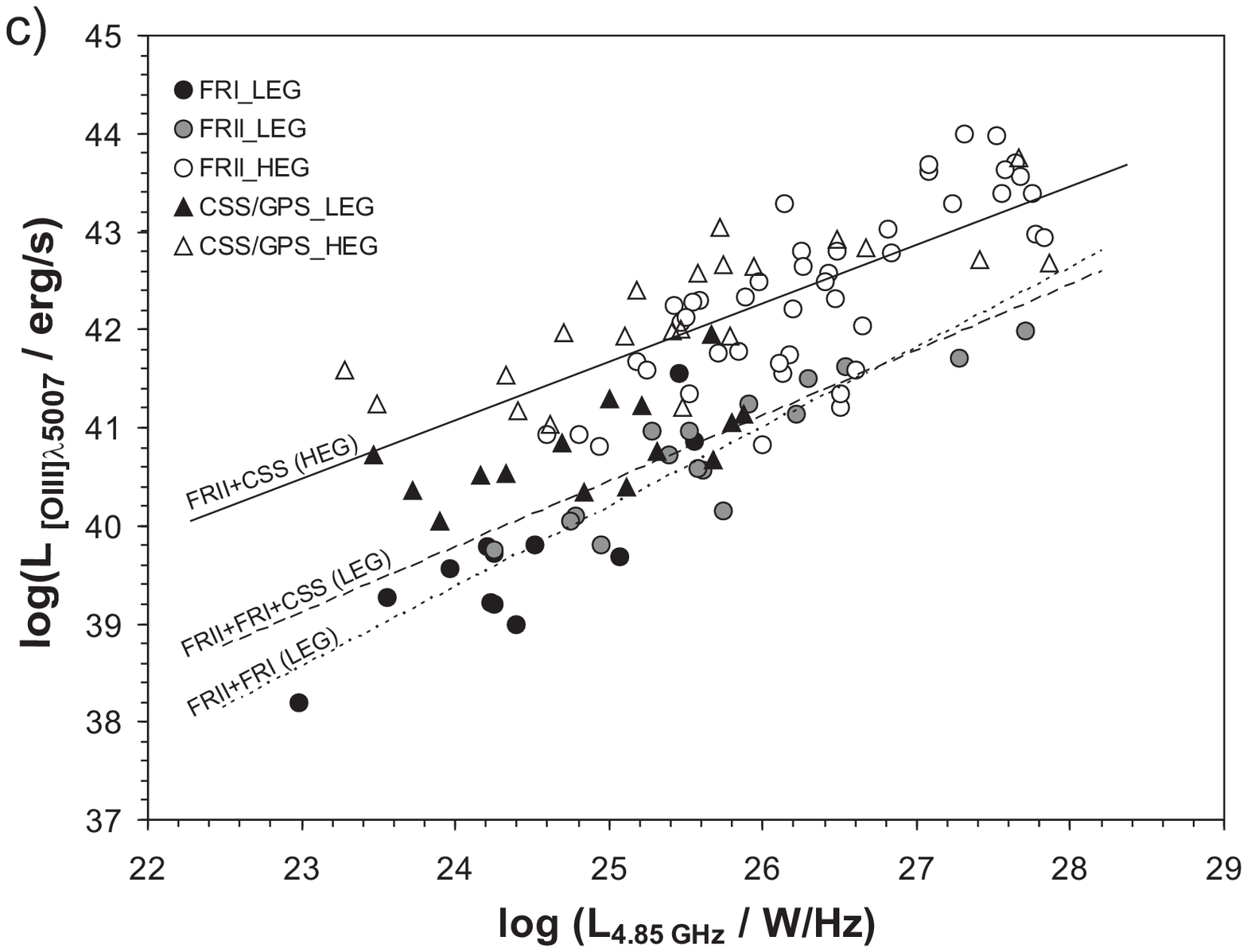}
\includegraphics[width=8.5cm,height=7cm]{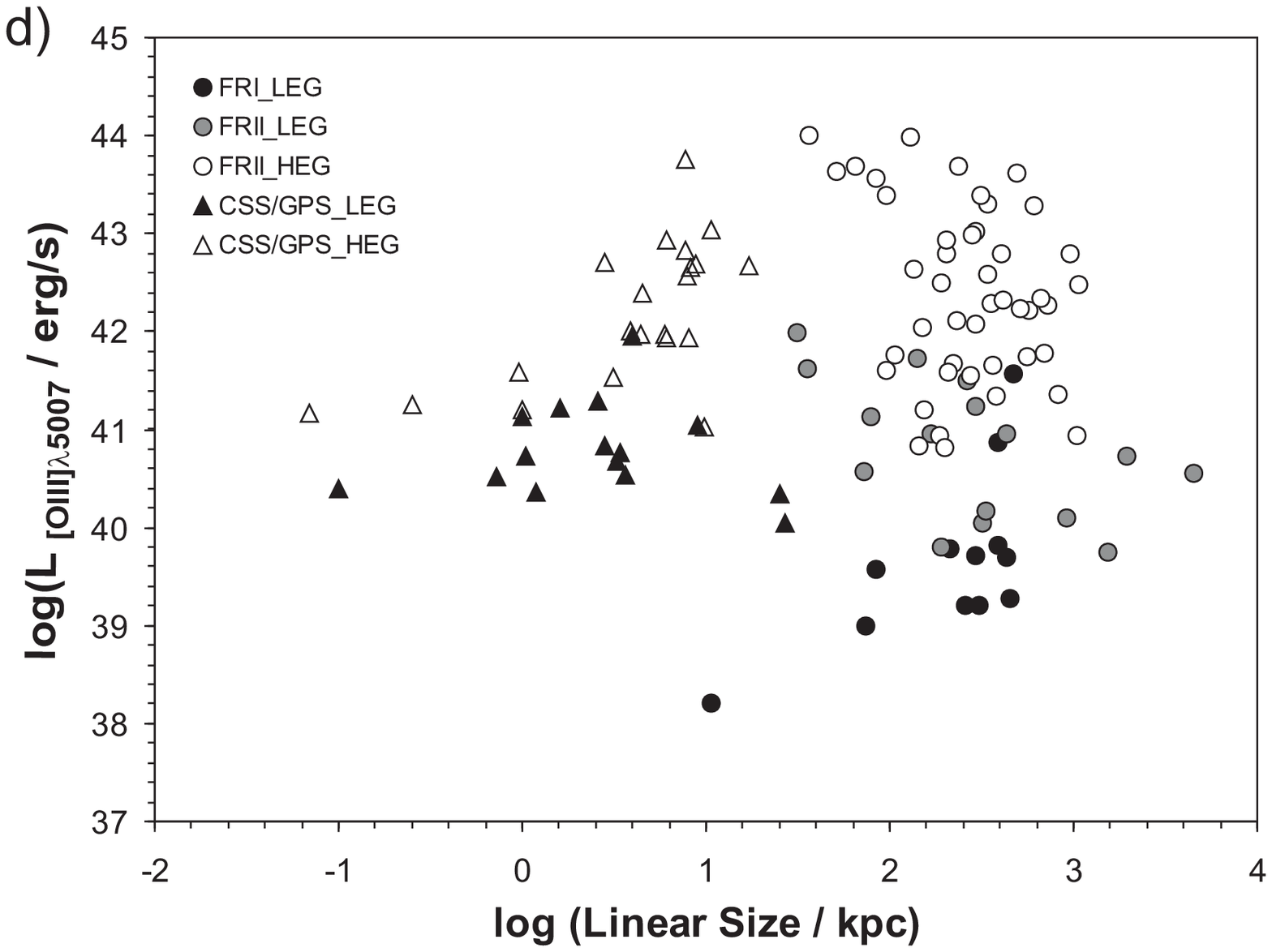}
\caption{a) [O\,III] luminosity-radio luminosity diagram and
b) [O\,III] luminosity-size diagram for AGNs; squares indicate CSS and GPS sources
from samples: \citet{labiano07}, LRL \citep{laing83, willott} and from
Table~\ref{combined} in this paper. The filled circles indicate FR\,I
objects and open circles indicate FR\,IIs from sample \citet{buti,willott}.
The triangles indicate LLC sources from this paper. 
c) [O\,III] luminosity-radio luminosity diagram and 
d) [O\,III] luminosity-size diagram for AGNs classified as HEG and LEG:
the samples used in these plots are taken from:
\citet{butti10,willott} (FR\,I/FR\,II, CSS), this paper and \citet{buti09} (CSS objects).
The values of optical and radio luminosities used in all plots are K-corrected. 
The linear size has been calculated based on the largest angular size
measured on resolved structure of the source taken from the literature.}
\label{OIIIandRadio}
\end{figure*}

\section{Discussion}
\label{dis}
Optical data are available for most of the radio sources. 
However, some data have too low signal-to-noise ($\la3$) and could not be analysed. 
We measured emission line fluxes of the LLC sources
(Table~\ref{lines}) and sources from \citet{f2001} sample and
\citet{mar03} sample for which available spectroscopic data exist
(Appendix, Table~\ref{combined}). 
Discussion on optical properties of CSS sources is then based on the results
presented in Fig. 1 and 2.
The radio properties of the sample have
been discussed and analysed in the Paper\,I.

\subsection{Spectroscopic properties of LLC sources}
\label{groups}

We have classified the LLC sources in HEG and LEG
following the criteria described in \citet{butti10} (Table~\ref{lines}).
However, not all sources in the sample had the necessary lines measured. 

It is worth mentioning that LLC HEGs show a $\sim10$ times higher [O\,III]
$\lambda$5007 luminosities than LLC LEGs. HEG show
L$_{[\rm O\,III]} > 10^{41.1}~{\rm erg~s^{-1}}$, while LEG L$_{[\rm O\,III]}
<
10^{41.1}~{\rm erg~s^{-1}}$.
This effect is not only present in our sample but also in \citet{butti10}.

Both HEG and LEG show similar numbers of quasars and galaxies. However,
all broad line objects (25\% of the LLC sample, 0923+079, 1053+505, 
1140+058, 1154+435, 1641+320) are quasars and have been classified as 
HEG. Large sources showing broad 
line emission (30\% of the 3CR sample) also fall under the HEG class.
They all show a broad H$_\beta$ component. 1154+435 shows broad $H_{\gamma}$ 
and H$_\alpha$ emission. 1641+320 shows broad $H_{\gamma}$ and $H_{\delta}$.
All broad line objects (and only these) 
have [O\,III] luminosity $>10^{42}~{\rm erg~s^{-1}}$.
The radio morphologies of 
1154+435 and 1641+320, are a core-jet structure with a bending jet and complex
four-component morphology respectively, indicating strong interaction with
the ISM (Paper\,I). An interesting case is a single-lobed object 0923+079,
candidate for a fader. While radio image (Paper\,I) suggests a fading
structure, the presence of broad emission lines could indicate that 
there is still interaction with the ISM.

Six of the LLC sources have low [O\,III] luminosity values ($<10^{41}~{\rm
erg~s^{-1}}$): 0810+077, 0851+024, 0921+143,
1007+142, 1037+302, 1558+536. These are both galaxies and quasars, all of them 
classified as LEG  Among them we have core-jet and asymmetric double objects. The radio
source 1037+302 has been classified as a young FR\,I \citep{gir05}, and 1558+536
with its breaking up radio structure is a candidate for a fader (Paper\,I).

The radio morphology properties of LLC HEG and LEG do not
correspond to a difference in their emission line properties indicating that
similar radio structures can be formed in both classes.

\subsection{Spectroscopic diagnostic diagrams}
\label{secbpt}

\begin{figure*}
\centering
\includegraphics[width=8cm,height=6cm]{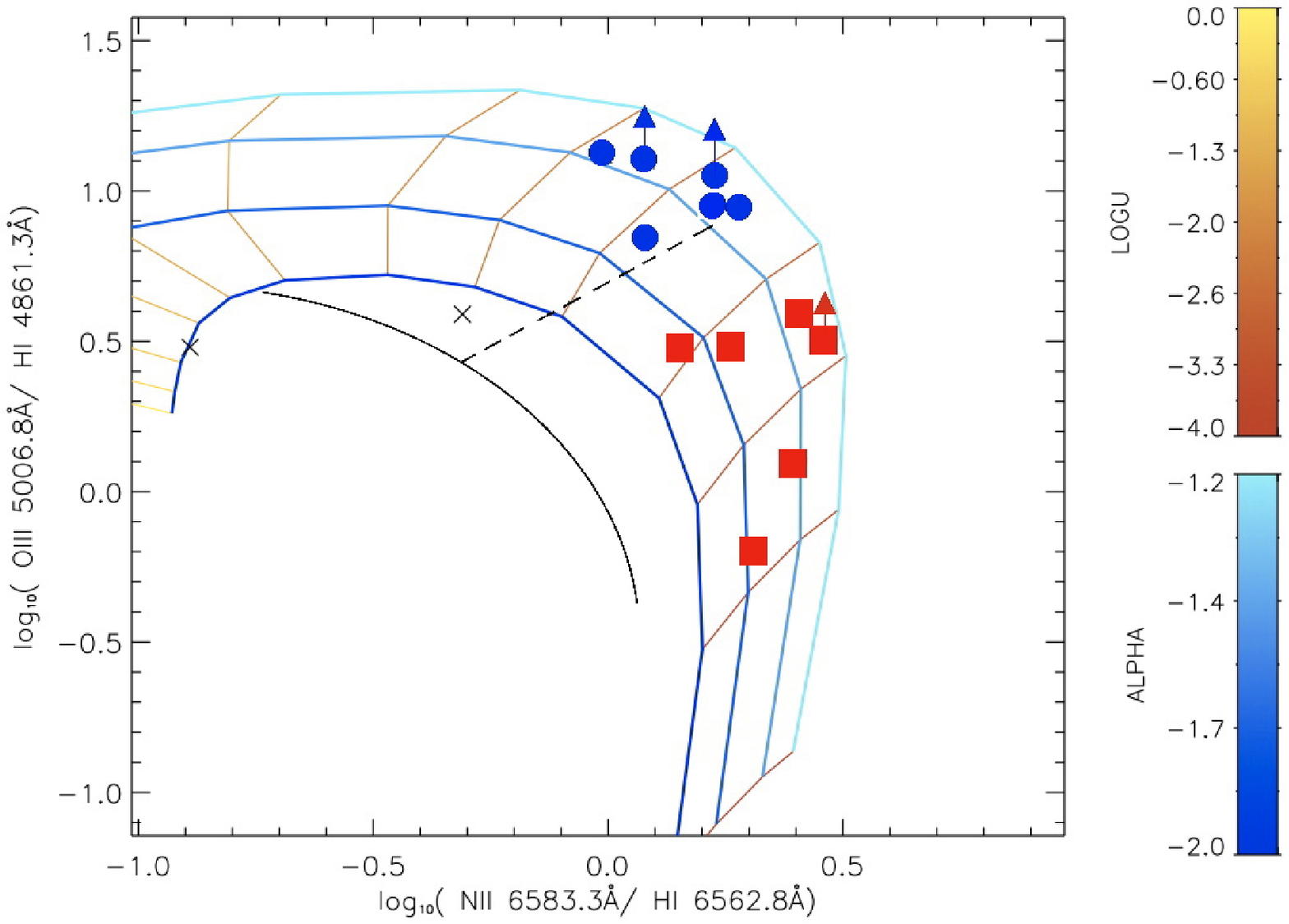} \hfil
\includegraphics[width=8cm,height=6cm]{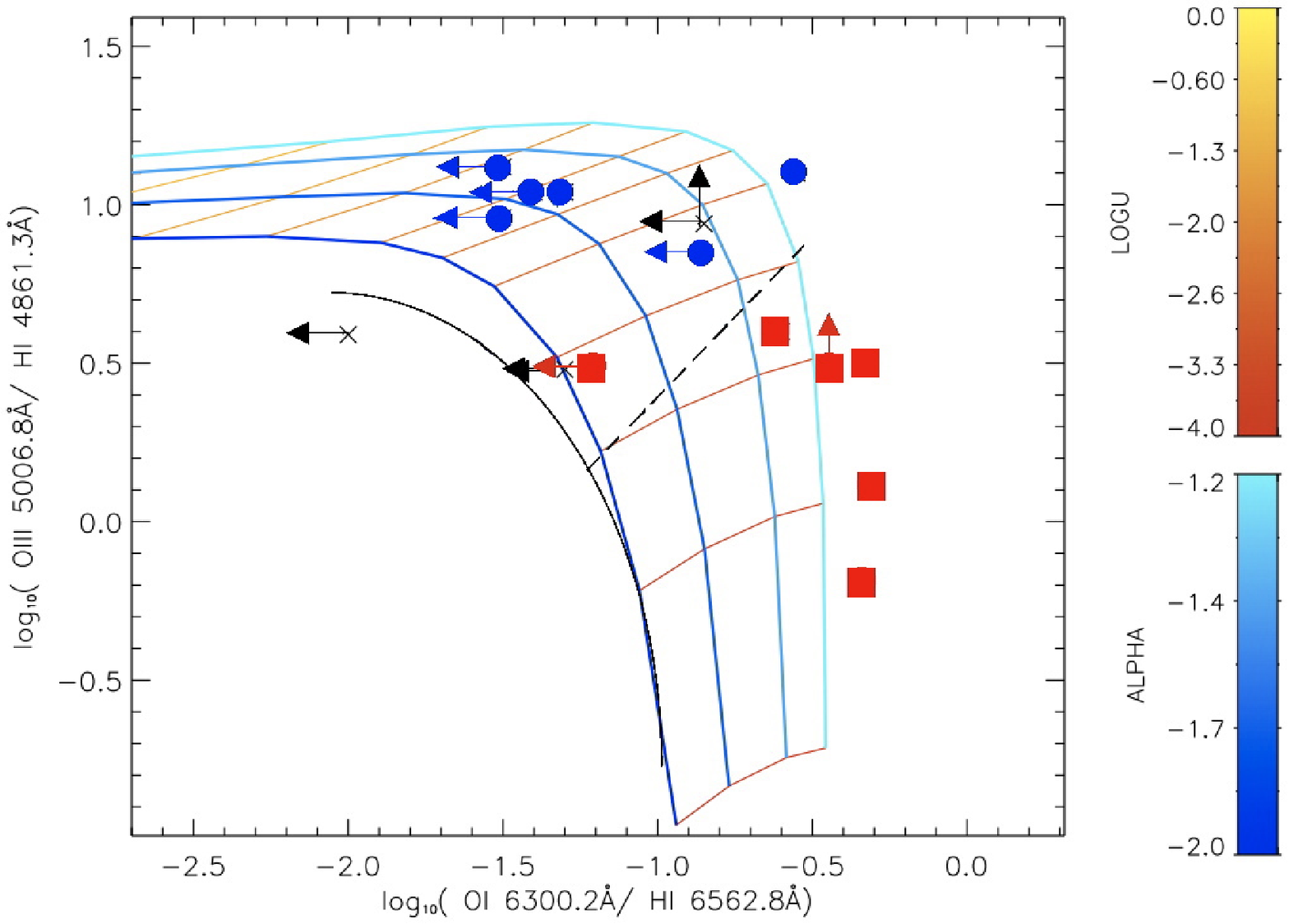} \hfil
\includegraphics[width=8cm,height=6cm]{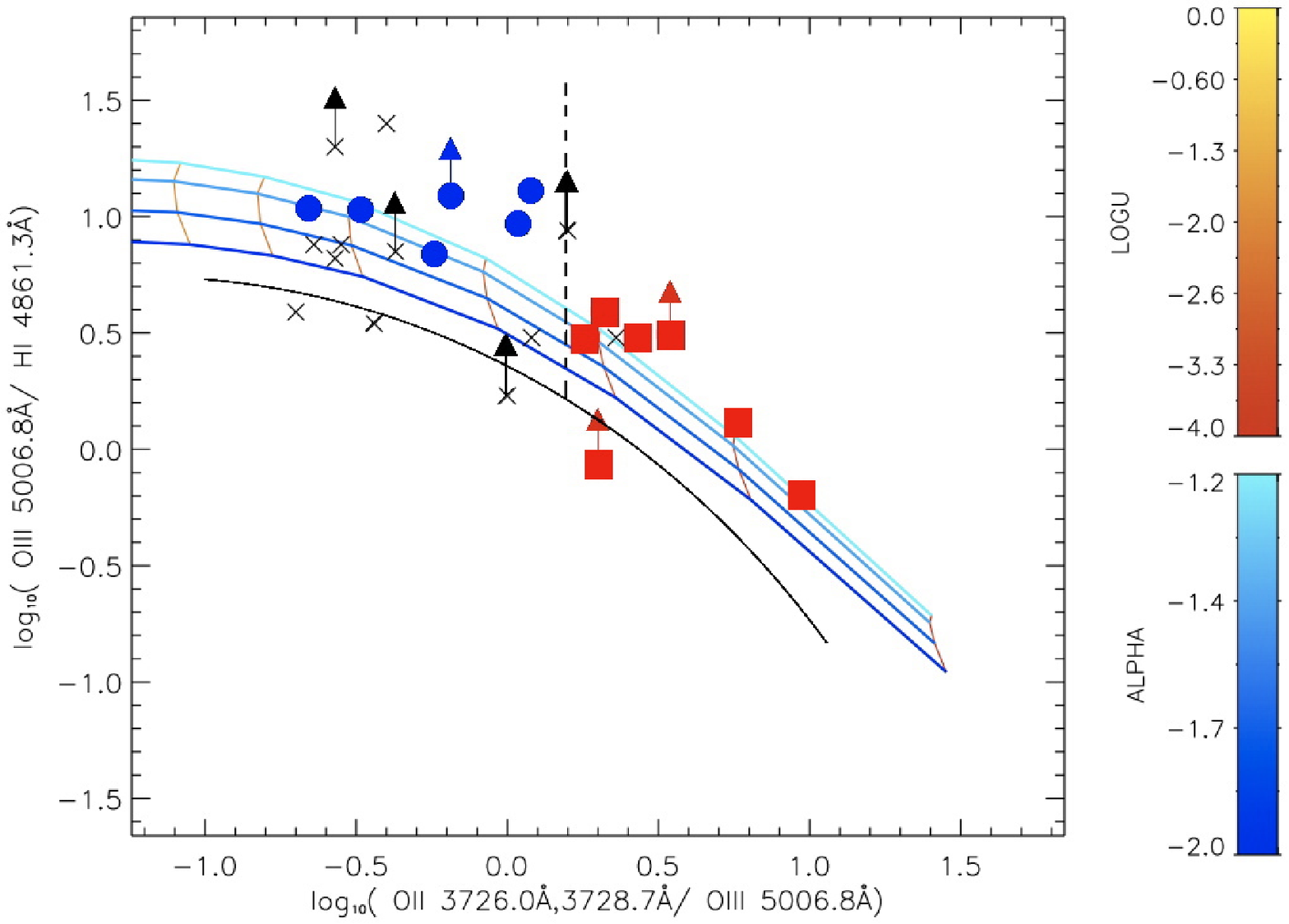} \hfil
\includegraphics[width=8cm,height=6cm]{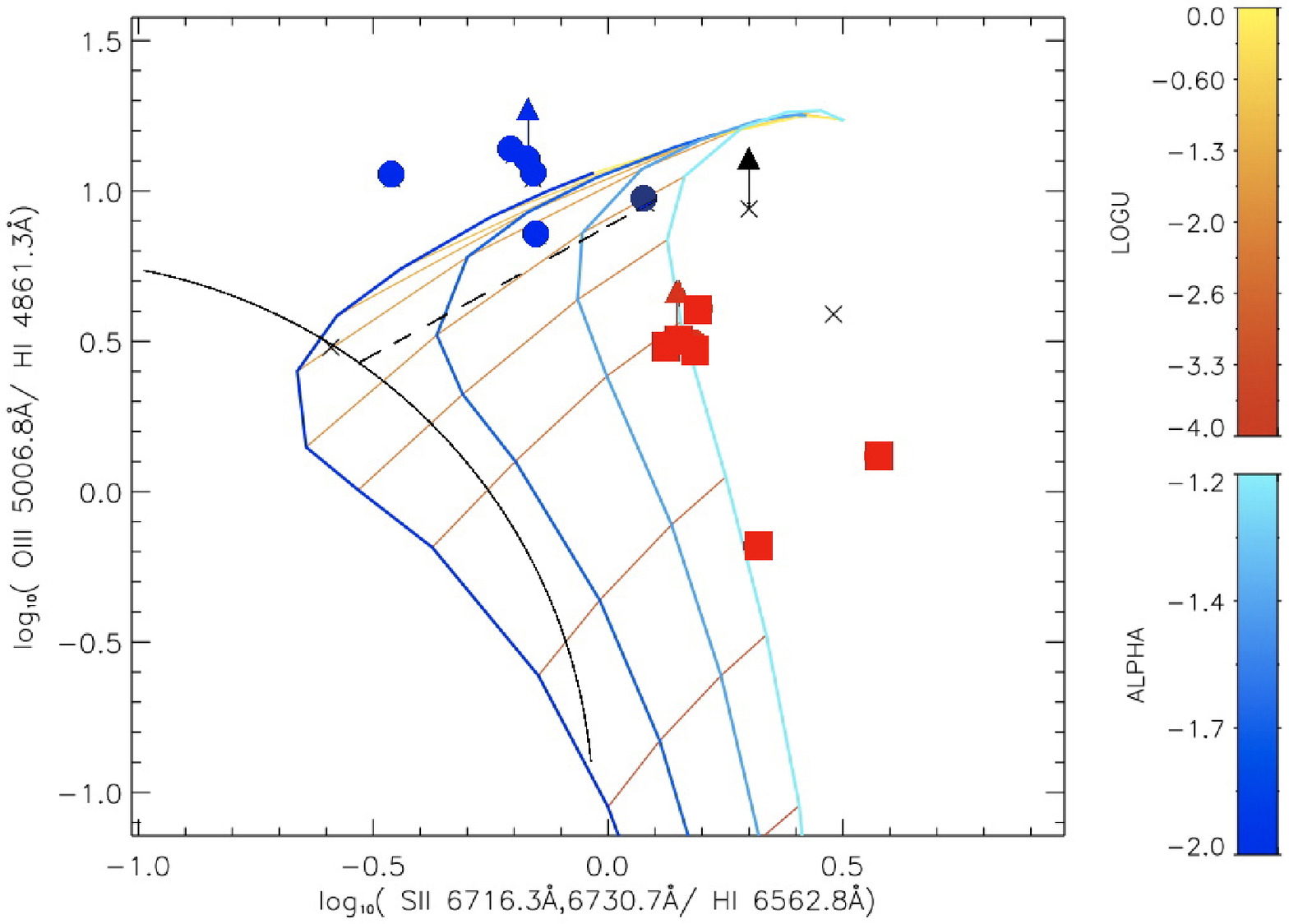} \hfil
\caption{Selected BPT diagrams for the sample of LLC sources. We show the best
fitting models to the sample. 
>From left to right and top to bottom: \nha\,dustfree AGN (Z=4, N=10000),
\oha\,dustfree AGN (Z=1, N=1000), \oo\,dustfree AGB (Z=1 N=1000), and
\sha\,dusty AGN (Z=2, N=1000). 
The remaining BPT diagrams and the colour version of this figure are available on-line. 
U - ionization parameter,
therefore the ratio of the photon density to the atomic density;
$\alpha$ - power-law index: F$\nu \propto \nu^\alpha$; Z - metallicity
(relative to Solar) of the ionized
gas in the model; N - hydrogen number density (cm$^{-3}$) of the modelled gas;
B=B'/n$^{1/2}$ - magnetic parameter
where {\it B'} is the transverse component of the preshock magnetic field
and
{\it n} is the preshock particle
number density. Velocity of the shock is in $\rm km~s^{-1}$. See \citet{Groves10} for 
details on the models. Red squares identify LEG, 
blue circles HEG, crosses unclassified sources. Solid lines mark the 
approximate division between AGN and HII regions \citep{Kewley06}, 
dashed lines mark the approximate division between HEG and LEG \citep{butti10}.}
\label{BPT}
\end{figure*}

We have used the ITERA \citep{Groves10} tool to to generate emission line diagnostic 
\citep[or BPT, after][]{baldwin} diagrams and compare the emission
line ratios (\oo, \ohb, \oha, \sha\,\-versus\,\- \nha) of the CSS HEG and LEG
sources in
our sample (Figure \ref{BPT}) with predictions of starburst \citep{Kewley01,
Dopita06, Levesque10},
dust, dustfree AGN \citep{Groves04a, Groves04b} and shock \citep{Allen08}
models.
Our results are roughly consistent with the 3CR sample studied by
\cite{butti10}.
However, our sample tend to show higher \nha, \sha\, and \oha\,  but
similar \ohb\, to the \cite{butti10} sample.

None of the recent ($<$ 10\,Myr) star formation models is consistent
with the data. The sources could be too weak to induce star formation or the gas has not
had time yet to cool down and form stars in the regions affected by the jet.

The higher ionization parameter and the presence of the precursor in HEG,
suggest a difference in the strength of the shocks and jet-ISM interactions
in HEGs and LEGs. Strong shocks (which show higher U and higher contribution from the
precursor gas, e.g.\citet{Allen08}) are present in HEGs while weaker shocks are
characteristic for LEGs.

Concerning the environment, both HEG and LEG are present at all gas densities and metalicities 
and are well reproduced by both dusty and dust-free AGN models. However,
HEGs tend to show slightly 
lower metallicities, and higher magnetic parameter 
\citep[ratio of the magnetic field to the density, see][for details]{Groves10}. 
The data also suggest stronger radiation fields in LEG environments although the measurements 
are not conclusive (most of the \oha \, data are limits).

\subsection{Emission lines - radio correlation}
\label{radio}

We have compared the [O\,III] luminosity with the radio properties for LLC sources, 
and expanded the sample with 
other CSS sources \citep{buti09,buti,butti10,willott}, GPS sources \citep{labiano07} and
FR\,I and FR\,II objects \citep{buti,willott}.

The whole sample shows that, for a given size or radio luminosity, HEG sources 
are brighter than LEG in the [O\,III] line (Table~\ref{lines}, 
Table~\ref{combined}, Fig.~\ref{OIIIandRadio}d) by a factor of $\sim$10.

The LLC objects follow the same correlation between [O\,III] luminosity and 
radio power, as the rest of the sample
(Figure ~\ref{OIIIandRadio}a). In Figure ~\ref{OIIIandRadio}b we plotted
the same sources in the plane [O\,III] luminosity versus linear size which
can be interpreted as a picture of radio source evolution. 
As we already showed and discussed in Paper\,I the LLC objects 
occupy the space in radio power versus
linear size diagram below the main evolutionary path of radio objects. 
Although we have optical data for 29 out of 44 LLC sources this trend is
also visible in Figure ~\ref{OIIIandRadio}b.
We
suggest that some of the low luminosity sources might be short-lived objects,
and their radio emission may be disrupted several times before becoming
FR\,IIs.  

The previously reported  correlation \citep{labianolet} between [O\,III]
luminosity and size of the radio source for CSS and GPS sources (although
not very strong, the correlation coefficient is 0.35) is breaking
up when including LLC objects (correlation coefficient equals to 0.11, Fig.~\ref{OIIIandRadio}b). 
It is also significant that CSS sources causing 
this effect are LEGs (Fig.~\ref{OIIIandRadio}d). This break up could be due to 
a sample selection effect since the \cite{labianolet} data was collected from the 
available literature samples which were created for different purposes 
and different selection criteria. The radio jet could only enhance the [O\,III] 
emission in HEGs, as these sources seem to be more powerful than LEGs.

Moreover, all sources (compact and large) classified as HEGs, have 
[O\,III] luminosities $\ga 10^{41}~{\rm erg~s^{-1}}$ while all LEGs fall below 
this limit and show stronger jet contributions in the BPT diagrams. 
(Fig.~\ref{OIIIandRadio}c,d).
It has been discussed \citep{butti10} that the differences between
LEGs and HEGs are related to a different mode of accretion: LEGs are powered by
hot gas while HEGs require the presence of cold accreting material.
The fact that only CSS HEGs seem to follow the correlation between [O\,III]
luminosity and size of the radio source for CSS/GPS sources \citep{labianolet}
can be connected with the above findings. According to \citet{labianolet}
the expansion of the radio source through the host ISM could be triggering or
enhancing the [O\,III]$\lambda5007$ line emission through direct
interaction. The presence of the cold structures (i.e. molecular torus or
Broad Line Region) in HEGs could be responsible for higher gas excitation, and
consequently for higher [O\,III]$\lambda5007$ line emission and $\rm
H{\alpha}$ line emission. Considering $\rm H{\alpha}$ line luminosities of
CSS sources we also found a division among them: CSS HEGs have
higher $\rm H{\alpha}$ line luminosities than CSS LEG objects 
(Table~\ref{lines} and Table~\ref{combined}) . 
This correlation has been earlier found for large scale FR\,Is and FR\,IIs
\citep{butti10}. 
However, in the case of CSS and GPS sources, more 
data are needed to confirm the correlation and study its possible implications.

A recent paper \citep{garofalo10}
suggest that the HEG/LEG nature of a radio galaxy could be related to the black 
hole spin instead of the different accretion mechanism \citep[e.g.][]{Hardcastle07}. 
However, the definitions of high and low excitation sources used in these two papers are 
slightly different than the definitions used by \cite{butti10} and adopted for our sample. 
While searching for HEG and LEG differences 
in the samples we used, we also found that they show differences in X-ray emission.
LEGs have X-ray 
luminosities below $10^{42}$ erg/s while HEG X-ray emission cover a wider range 
\citep[from $0.1\times 10^{42}$ to $1.4\times10^{44}$, e.g.][]{Evans06, Hardcastle06, Tengstrand09, Massaro10}.
Although there is an overlap around $10^{41-42}$, there are no LEGs with X-ray luminosities higher 
than 10$^{42}$ erg/s. We see the same behaviour in [OIII]$\lambda50007$ luminosity (with a similar 
overlap around 10$^{41}$erg/s, see Figure 1 c and d). A detailed study about the origin of the HEG and 
LEG X-ray properties is however beyond the scope of this paper. 

Figure~\ref{OIIIandRadio}c shows that the division for HEGs and LEGs described 
for FR\,II/FR\,I sources by \citet{butti10} is also visible among CSS sources 
(i.e. earlier phases of AGN evolution). 
The FR\,II HEG and CSS HEG, as well as the FR\,II LEG and CSS LEG show a wide 
range of radio powers, but it seems they follow different [O\,III] luminosity 
versus radio power correlation. FR\,I sources could be completely
different group of objects concerning the photoionization, or follow the LEG
[O\,III] luminosity versus radio power correlation. As has been noted by
\citet{butti10} all FR\,Is for which they were able to derive a spectral
type are LEGs.
We considered separately the populations of HEGs and LEGs and we have obtained the 
following linear correlations (plotted in Fig.~\ref{OIIIandRadio}c):

\noindent
HEG (FR\,II+CSS):\\
$\log{L_{[\rm O\,III]}}=0.59(\pm 0.07)\times\log{L_{\rm 4.85\,GHz}}+26.77(\pm
1.75)$\\

\noindent
LEG (FR\,I+FR\,II+CSS):\\
$\log{L_{[\rm O\,III]}}=0.67(\pm 0.08)\times\log{L_{\rm
4.85\,GHz}}+23.69(\pm 2.12)$\\

\noindent
LEG (FR\,I+FR\,II):\\
$\log{L_{[\rm O\,III]}}=0.81(\pm 0.07)\times\log{L_{\rm
4.85\,GHz}}+19.79(\pm 1.76)$\\

Based on the analysis above, we propose a scenario where the differences in the nature 
of LEG and HEG (accretion mode or black hole spin) are already visible in the CSS
phase of AGN and determine the evolution of the source (i.e. $\rm CSS_{LEG}$ evolve to
$\rm FR_{LEG}$, $\rm CSS_{HEG}$ evolve to $\rm FR_{HEG}$. The main evolution scenario 
(GPS-CSS-FR\,II) was proposed years ago \citep{f95, r96}. 
However, once the HEG/LEG division is included, these sources seem to evolve in parallel: 
${\rm GPS_{LEG}-CSS_{LEG}-FR_{LEG}}$ and
${\rm GPS_{HEG}-CSS_{HEG}-FR\,II_{HEG}}$. 
Concerning LEG, it is still not clear 
if $\rm CSS_{LEG}$ would evolve directly to $\rm FR\,I_{LEG}$ or go through a 
$\rm FR\,II_{LEG}$ phase before the $\rm FR\,I_{LEG}$.

As discussed in Paper\,I there should also exist a group
of short-lived CSS objects with lower radio luminosities. 
These short-lived CSSs could probably show the low [O\,III] 
luminosities seen in FR\,Is.

The nature of the division for HEGs and LEGs among radio galaxies is still a
debated issue and waiting for an answer. It seems that the radio and ionized
gas luminosities of radio galaxies are determined by the properties of central 
engine: black hole spin \citep{garofalo10}, rate \citep{willott} or mode of accretion
\citep{Hardcastle07,butti10}. In the last case the authors working on the
large scale FR\,IIs and FR\,Is speculate that HEGs are powered by accretion of
cold gas (provided probably by the merger with a gas rich galaxy), while
LEGs
accreate hot material. As we have just shown, the HEG/LEG
division is present also among the young radio galaxies: CSS and GPS
galaxies.
We speculate that the differences in central engine of radio galaxy can occur
at the beginning of the activity phase. However, as we discuss in
Paper\,I, many of the low luminosity CSSs can be short-lived objects and undergo the
CSS phase of evolution many times, as they are able to escape from the
host galaxy and evolve further. The ignition of the activity can be caused
by different mechanisms, major or minor mergers or instabilities in the
accretion flow \citep{czerny}. Depending on the ignition mechanism the
radio galaxy can return as LEG or HEG. The comparison of the properties of LLC
sources with different ionization models suggest also 
that jet-ISM interactions play an important role on producing HEG (through
strong shocks in the ISM) or LEG excitation levels (weaker shocks).
However, spectroscopic observations of larger sample of CSS and GPS sources
and also studying their environmental properties are needed to give more
certain conclusions.

\section{Summary}

In this paper, we present and discuss the optical properties of Low Luminosity 
Compact sources, based on SDSS images and spectra. The sample and
its radio properties are presented in Paper\,I. 
The comparison of the LLC sources with luminous CSSs and large radio sources 
has given a wider view of how LLC objects fit to the general radio source scenario.

Using the emission line ratios, we 
classified the LLC sources as HEGs and LEGs. 
For the same size and radio luminosity, HEGs are 10 times more luminous in [O III] than
LEGs. 
This behaviour is also present in brighter CSS and large radio sources. Furthermore, 
LLC HEGs and LEGs fall above and below L$_{[\rm O\,III]}=10 ^{41.1}$ erg/s, respectively. 
All sources with with broad line emission in their spectra are HEG. They are also the most 
luminous objects in our sample (L$_{[\rm O\,III]} > 10^{42}$ erg/s).

We have compared the [O\,III] luminosity with radio power linear size
for the LLC sample, bright CSS and large radio sources.
The correlation reported for bright GPS/CSS and large sources, between
[O\,III] luminosity and radio size, disappears when LLC objects 
are included, mainly because to CSS LEGs. This effect could be 
to result of a previous sample selection effect. However, it is also possible 
that the radio jet could only enhance the [O\,III] emission in HEG, which 
show a stronger jet contribution to the ionization of the ISM than LEG.

The distribution of the LLC sample in the plots is consistent with the 
correlations found for CSS and large radio sources, although LLC sources 
have lower [O\,III] luminosities.
However, when the samples are separated into HEGs and LEGs, they seem to follow 
different correlations. Furthermore, the HEG/LEG classification in the LLC,
bright CSS and large-scale sources is independent of radio power and size. 
Concerning the radio morphology, CSS behave like FR\,II sources at all
radio powers ,suggesting that 
the differences in the mode of accretion (or black hole spin) between 
LEG and HEG sources is already visible in the CSS phase. 
According to BPT diagrams generated for LLC sources, HEGs have stronger
shocks than LEGs, which can indicate differences in their environment. 

Based on these results, we suggest that HEG and LEG follow different 
(and parallel) evolutionary tracks during the whole life 
of the radio source: ${\rm GPS_{LEG}-CSS_{LEG}-FR_{LEG}}$ and 
${\rm GPS_{HEG}-CSS_{HEG}-FR_{HEG}}$. We could still 
be missing the precursors of the large FR\,I sources. However, 
short-lived CSS could probably show the low [O\,III] 
luminosities seen in FR\,I and give some clues as to their origin.

\section*{Acknowledgments}
We thank Drs. Matteo Guainazzi, Sara Buttiglione and Philip Best for many useful comments and 
fruitful discussions. 

This research has made use of NASA's Astrophysics Data System (ADS) and 
the NASA/IPAC Extragalactic Database (NED) which is operated by the Jet Propulsion Laboratory, 
California Institute of Technology, under contract with the National Aeronautics and 
Space Administration. Funding for the SDSS and SDSS-II has been provided by the Alfred 
P. Sloan Foundation, the Participating Institutions, the National Science Foundation, the 
U.S. Department of Energy, the National Aeronautics and Space Administration, the Japanese 
Monbukagakusho, the Max Planck Society, and the Higher Education Funding Council for England. 
The SDSS Web Site is http://www.sdss.org/. The SDSS is managed by the Astrophysical Research 
Consortium for the Participating Institutions. The Participating Institutions are the American 
Museum of Natural History, Astrophysical Institute Potsdam, University of Basel, University of 
Cambridge, Case Western Reserve University, University of Chicago, Drexel University, Fermilab, 
the Institute for Advanced Study, the Japan Participation Group, Johns Hopkins University, 
the Joint Institute for Nuclear Astrophysics, the Kavli Institute for Particle Astrophysics 
and Cosmology, the Korean Scientist Group, the Chinese Academy of Sciences (LAMOST), Los 
Alamos National Laboratory, the Max-Planck-Institute for  Astronomy (MPIA), 
the Max-Planck-Institute for Astrophysics (MPA), New Mexico State University, Ohio State 
University, University of Pittsburgh, University of Portsmouth, Princeton University, the 
United States Naval Observatory, and the University of Washington. 

This work makes use of EURO-VO software, tools or services. The EURO-VO has been funded by 
the European Commission through contract numbers RI031675 (DCA) and 011892 (VO-TECH) under 
the 6th Framework Programme and contract number 212104 (AIDA) under the 7th Framework Programme.)

\appendix

\section[]{Combined sample of CSS sources}

The spectroscopic data are available for 14 CSS sources from the
combined sample (Table A1). The HEG/LEG classification was possible for only
3 of them. 

\noindent {\bf Notes on individual sources}

\noindent {\bf 0800+472}.
HEG*. The image shows a compact source ($\sim$16\,kpc) with a possible
eastern companion $\sim$50\,kpc.
The spectrum shows a powerlaw continuum and broad [Mg\,II]$\lambda$2800
emission, consistent with a QSO.
The radio map shows compact source \citep{f2001}, consistent with the QSO
optical
classification.

\noindent {\bf 0809+404}.
HEG*. The image shows a compact source ($\sim$12\,kpc) with a possible tail
towards W, consistent with the faint component in the radio map
\citep{f2001}.

\noindent {\bf 1141+466}.
The image shows an extended source ($\sim$20\,kpc) with a bright nucleus and
bright emission
knots in a very crowded field.

\noindent {\bf 1201+394}.
The image shows a compact source ($\sim$16\,kpc), slightly elongated towards
NW, consistent with the orientation of the radio structure \citep{f2001}.   

\noindent {\bf 1241+411}.
HEG. The image shows an extended source ($\sim$40\,kpc) with a bright
nucleus.

\noindent {\bf 1343+386}.
The optical data show a bright compact source ($\sim$20\,kpc) with a clear
powerlaw and bright broad emission lines, suggesting a QSO.

\noindent {\bf 1445+410}.
The image shows an extended source with a bright nucleus. There is no clear
evidence of the $\sim$20\,kpc radio structure \citep{f2001} in the optical
image.

\noindent {\bf 0801+303}.
The optical data show a bright compact source ($\sim$30\,kpc) with a clear
powerlaw and bright broad emission lines, suggesting a QSO.

\noindent {\bf 0853+291}.
The optical data show a bright compact source ($\sim$30\,kpc) with a clear
powerlaw and bright
broad emission lines, suggesting a QSO.

\noindent {\bf 1251+308}.
The optical data show a bright compact source ($\sim$30\,kpc) with a clear
powerlaw and bright
broad [Mg\,II]$\lambda$2800 emission, suggesting a QSO. The radio map shows
a small
source with a very
structured $\sim$5\,kpc long jet oriented SE-NW \citep{mar06}.

\noindent {\bf 1315+396}.
The optical data show a bright compact source ($\sim$30\,kpc) with a clear
powerlaw and bright
broad emission lines, suggesting a QSO. The radio map shows a very small
($\sim$0.03\,kpc) jet oriented EW \citep{kun06}.

\noindent {\bf 1502+291}.
The optical data show a bright compact source ($\sim$30\,kpc) with a clear
powerlaw and bright
broad emission lines, suggesting a QSO.

\noindent {\bf 1619+378}.
The image shows a compact faint source suggesting a possible QSO. The
spectrum shows only one
clear line, with broad wings. If it is [Mg\,II]$\lambda$2800, then the
source
is at redshift $z=1.2734$.
However, this line could be affected by a possible break in the continuum.
It
also shows no clear
traces of the expected AGN powerlaw. The radio map \citep{mar06} shows a
0.5$\arcsec$
($\sim$4\,kpc at z=1.2734)
jet oriented NE-SW, which does not show up in the optical image.

\noindent {\bf 1632+391}.
The spectrum shows a clear powerlaw with bright broad emission lines,
consistent with a QSO.
The image shows a two component source, separated $\sim$30\,kpc. The W
component is compact and
bright while the E component is fainter and extended. The E component was
classified as a compact
blue cluster by \citet{hut95}. The radio map \citep{mar06} shows a
$\sim$6\,kpc jet
oriented NE-SW at the
coordinates of the compact source, which does not show up in the optical
image.

Summing up, 9 sources (64\%) are compact or unresolved, 5 sources (36\%)
show extended emission,
and 1 source (7\%) shows a possible merger or complex structure. Only 2
sources (14\%) show 
radio-optical alignment.

\begin{table*}
\begin{center}
\caption[]{Emission lines measurements and spectroscopic classification of
the combined sample of CSS sources - part I.}
\begin{tabular}{@{}c c c c c c c c c c c @{}}
\hline
Source & Class&
\multicolumn{1}{c}{\it z}&
\multicolumn{1}{c}{\it E(B-V)}&
\multicolumn{1}{c}{[Ly$\alpha$] } &
\multicolumn{1}{c}{[C IV]} &
\multicolumn{1}{c}{[He II]} &
\multicolumn{1}{c}{[C III]} &
\multicolumn{1}{c}{[Mg II]} &
\multicolumn{1}{c}{[O II]} & 
\multicolumn{1}{c}{[Ne III]} \\ 
Name	& & & & &
\multicolumn{1}{c}{$\lambda$ 1549} &
\multicolumn{1}{c}{$\lambda$ 1640}&
\multicolumn{1}{c}{$\lambda$ 1908}&
\multicolumn{1}{c}{$\lambda$ 2800}&
\multicolumn{1}{c}{$\lambda$ 3727,3729} &
\multicolumn{1}{c}{$\lambda$ 3870} \\
\hline
0800+472& HEG*	&	0.51	&	0.04	&	--	&	--
&	--	&	--	&	48	&	22	&	8\\
0809+404&HEG*	&	0.55	&	0.05	&	--	&	--
&	--	&	--	&	--	&	16	&	--\\
1141+466& --	&	0.12	&	0.02	&	--	&	--
&	--	&	--	&	--	&	32	&	--\\
1201+394& --	&	0.45	&	0.03	&	--	&	--
&	--	&	--	&	--	&	2	&	--\\
1241+411&HEG	&	0.25	&	0.02	&	--	&	--
&	--	&	--	&	--	&	18	&	4\\
1343+386& --	&	1.85	&	0.01	&	--	&	67
&	--	&	32	&	36	&	--	&	--\\
1445+410& --	&	0.2	&	0.02	&	--	&	--
&	--	&	--	&	--	&	3	&	--\\
\hline
0801+303& --	&	1.45	&	0.04	&	--	&	--
&	3	&	42	&	28	&	3	&	--\\
0853+291& --	&	1.09	&	0.03	&	--	&	--
&	--	&	28	&	18	&	18	&	4\\
1251+308& --	&	1.31	&	0.02	&	--	&	--
&	--	&	--	&	5	&	--	&	--\\
1315+396& --	&	1.56	&	0.01	&	--	&	48
&	--	&	38	&	26	&	--	&	--\\
1502+291& --	&	2.28$\dagger$	&	0.02	&	87	&	41
&	--	&	12	&	8	&	--	&	--\\
1619+378& --	&	1.27	&	0.01	&	--	&	--
&	--	&	--	&	4	&	--	&	--\\
1632+391& --	&	1.09	&	0.01	&	--	&	--
&	--	&	21	&	14	&	2	&	--\\
\hline
Avg. Error& -- & -- & -- &8\%	&	7\%	&	30\%	&	12\%	&	19\%
&	8\%	&	8\%	\\

\hline
\end{tabular}
\end{center}
\end{table*}

\setcounter{table}{0}
\begin{table*}
\begin{center}
\caption[]{Emission lines measurements and spectroscopic classification of
the combined sample of CSS sources - part II}
\begin{tabular}{@{}c c c c c c c c c c c c @{}}
\hline
Source & 
\multicolumn{1}{c}{H$_\delta$} &
\multicolumn{1}{c}{H$_\gamma$} &
\multicolumn{1}{c}{H$_\beta$} &
\multicolumn{1}{c}{[O III]} &
\multicolumn{1}{c}{[O III]} &
\multicolumn{1}{c}{[O I]} &
\multicolumn{1}{c}{[N II]} &
\multicolumn{1}{c}{H$_\alpha$} &
\multicolumn{1}{c}{[N II]} &
\multicolumn{1}{c}{[S II]} &
\multicolumn{1}{c}{[S II]} \\

Name	& & & &
\multicolumn{1}{c}{$\lambda$ 4959} &
\multicolumn{1}{c}{$\lambda$ 5007} &
\multicolumn{1}{c}{$\lambda$ 6300} &
\multicolumn{1}{c}{$\lambda$ 6548} & &
\multicolumn{1}{c}{$\lambda$ 6584} &
\multicolumn{1}{c}{$\lambda$ 6716} &
\multicolumn{1}{c}{$\lambda$ 6731} \\
\hline
0800+472& 2& --& --	&28	&	85
&	--	&	--	&	--	&	--	&	--
&	--	\\
0809+404& --& 2& 7     &19	&	57
&	--	&	--	&	--	&	--	&	--
&	--	\\
1141+466& --& 4& 4	&1	&	--
&	--	&	14	&	22	&	41	&	15
&	9	\\
1201+394& --&--& --     &--	&	2
&	--	&	--	&	--	&	--	&	--
&	--	\\
1241+411& --&--& 5	&19	&	55
&	--	&	37	&	32	&	53	&	15
&	7	\\
1343+386& --&--&--&	--	&	--
&	--	&	--	&	--	&	--	&	--
&	--	\\
1445+410& --&--&--&	--	&	--
&	3	&		&	18	&	--	&	--
&	--	\\
\hline
0801+303& --&--&--&	--	&	--
&	--	&	--	&	--	&	--	&	--
&	--	\\
0853+291& --&6&--&	--	&	--
&	--	&	--	&	--	&	--	&	--
&	--	\\
1251+308& --&--&--&	--	&	--
&	--	&	--	&	--	&	--	&	--
&	--	\\
1315+396& --&--&--&	--	&	--
&	--	&	--	&	--	&	--	&	--
&	--	\\
1502+291& --&--&--&	--	&	--
&	--	&	--	&	--	&	--	&	--
&	--	\\
1619+378& --&--&--&	--	&	--
&	--	&	--	&	--	&	--	&	--
&	--	\\
1632+391& --&--&--&	--	&	--
&	--	&	--	&	--	&	--	&	--
&	--	\\
\hline
Avg. Error & 31\%& 44\% & 19\%&	15\% &	7\% &	13\%	&	24\%	&	16\%	&	11\%	&	12\% &	17\%	\\
\hline
\end{tabular}
\end{center}
\label{combined}

\begin{minipage}{160mm}
{\small
Description of the columns:
(1) source name;
(2) spectroscopic classification, '*' means the
classification is based on the \oo \, ratio only;
(3) redshift, '$\dagger$' - this is a new value from the SDSS; 
(4) galactic extinction;
Rest of columns: flux for each line in $10^{-16}~erg~s^{-1}~cm^{-2}$.
Spectral information are taken from SDSS.
Sources are taken from \citet{f2001} sample (first part of the table) and from
\citet{mar03}
sample. There is a small overlap of this two sample that is why the
overlapping sources are included only in a first part of the table.
}
\end{minipage}
\end{table*}


\begin{thebibliography}{}

\bibitem[Abazajian et al.(2009)]{Abazajian} Abazajian et al., 2009, ApJS,
182, 543

\bibitem[Allen et al.(2002)]{allen} Allen, M. G., Sparks, W. B.,
Koekemoer, A., et al., 2002, ApJS, 139, 411

\bibitem[Allen et al.(2008)]{Allen08} Allen, M.~G., Groves, B.~A., 
Dopita, M.~A., Sutherland, R.~S., Kewley, L.~J., 2008, ApJS, 178, 20.

\bibitem[Axon et al.(2000)]{Axon00} Axon, D.~J., Capetti, A., Fanti, R., Morganti,
R., Robinson, A., Spencer, R., 2000, AJ, 120, 2284

\bibitem[Baldwin et al.(1981)]{baldwin} Baldwin, J. A., Phillips, M. M.,
Terlevich, R., 1981, PASP, 93, 5

\bibitem[Brotherton et al.(1999)]{broth} Brotherton, M.~S., Gregg, M.~D.,
Becker, R.~H., Laurent-Muehleisen, S.~A., White, R.~L., Stanford, S.~A.,
1999, ApJ, 514, L61

\bibitem[Buttiglione et al.(2010)]{butti10} Buttiglione S., Capetti, A.,
Celotti, A., Axon, D.J., Chiaberge, M., Macchetto, F.D., Sparks, W.B., 2010,
A\&A, 509, 6

\bibitem[Buttiglione et al.(2009b)]{buti} Buttiglione S., Capetti, A.,
Celotti, A., Axon, D.J., Chiaberge, M., Macchetto, F.D., Sparks, W.B., 2009b,
A\&A, 495, 1033

\bibitem[Buttiglione et al.(2009a)]{buti09} Buttiglione, S.; Celotti, A.; Capetti, A.; 
Dallacasa, D.; D'Odorico, V.; Giovannini, G., 2009a, AN, 330, 237

\bibitem[Cardelli et al.(1989)]{cardelli} Cardelli, J. A., Clayton, G. C.,
Mathis, J.
S., 1989, ApJ, 345, 245

\bibitem[Chambers et al.(1987)]{chambers87} Chambers, K.~C.; Miley, G.~K.; 
van Breugel, W, 1987, Nature, 329, 604

\bibitem[Chiaberge et al.(2002)]{chiaberge} Chiaberge, M., Macchetto,
F. D., Sparks, W. B., et al. 2002, ApJ, 571, 247

\bibitem[Czerny et al.(2009)]{czerny} Czerny, B., Siemiginowska, A., Janiuk,
A., Nikiel-Wroczy\'nski, B., Stawarz, {\L.}, 2009, ApJ, 698, 840

\bibitem[Dopita et al.(2006)]{Dopita06} Dopita, M.~A., Fischera, J., Sutherland,
R.~S., Kewley, L.~J., Leitherer, C., Tuffs, R.~J., Popescu, C.~C., 
van Breugel, W., Groves, B.~A., 2006, ApJS, 167, 177.

\bibitem[Evans et al. (2006)]{Evans06} Evans, D.~A., Worrall, D.~M.,
Hardcastle, M.~J., Kraft, R.~P., Birkinshaw, M., 2006, ApJ, 642, 96

%\bibitem[Fanaroff \&\-Riley(1974)]{fr74} Fanaroff, B.~L., \& Riley,
%J.~M. 1974, MNRAS 167, 31

\bibitem[Fanti et al.(1995)]{f95} Fanti, C., Fanti, R., Dallacasa, D., et
al. 1995, A\&A 302, 317

\bibitem[Fanti et al.(2001)]{f2001} Fanti, C., Pozzi, F., \& Dallacasa, D.,
et~al. 2001, A\&A 369, 380

\bibitem[Garofalo et al. (2010)]{garofalo10} Garofalo, D., Evans, D. ~A. 
\& Sambrna, R.~M., 2010, arXiv:1004.1166.

\bibitem[Giroletti et al.(2005)]{gir05} Giroletti, M., Giovannini, G., \&
Taylor, G.~B. 2005, A\&A 441, 89

\bibitem[Groves \& Allen(2010)]{Groves10} Groves, B.~A \& Allen, M.~G, 2010, arxiv:1002.3372

\bibitem[Groves et al.(2004a)]{Groves04a} Groves, B.~A., Dopita, M.~A., Sutherland, R.~S., 2004, ApJS, 153, 9

\bibitem[Groves et al.(2004b)]{Groves04b} Groves, B.~A., Dopita, M.~A., Sutherland, R.~S., 2004, ApJS, 153, 75

\bibitem[Hardcastle et al.(2007)]{Hardcastle07} Hardcastle, M. J.; 
Evans, D. A.; Croston, J. H., 2007, MNRAS, 376, 1849

\bibitem[Hardcastle et al.(2006)]{Hardcastle06} Hardcastle, M. J.; 
Evans, D. A.; Croston, J. H., 2006, MNRAS, 370, 1893

\bibitem[Holt et al.(2009)]{holt09} Holt, J., Tadhunter, C.~N., \& Morganti,
R., 2009, MNRAS, 400, 589

\bibitem[Holt et al.(2006)]{holt} Holt, J., Tadhunter, C.~N., \& Morganti,
 R., 2006, AN, 327, 147

\bibitem[Hutchings et al.(1995)]{hut95} Hutchings, J.~B., Crampton, D.,
Johnson, A., 1995, AJ, 109, 73

\bibitem[Kai\-ser\- \& Best(2007)]{kb07} Kaiser, C.~R., \& Best, P.~N.
2007, MNRAS 381, 1548

\bibitem[Kewley et al. (2001)]{Kewley01} Kewley, L.~J., Dopita, M.~A., Sutherland, R.~S., Heisler, C.~A., Trevena, J., 2001, ApJ, 556, 121

\bibitem[Kewley et al. (2006)]{Kewley06} Kewley, L.~J., Groves, B.; Kauffmann, G. Heckman, T. 2006, AJ, 372, 961

\bibitem[Kopylov et al.(1995)]{koylov95} Kopylov A. I., Goss, V. M.,
Pariiskii, Yu. N., Soboleva, N. S., Zhelenkova, O. P., Tempirova, A. V.,
Vitkovskii, V. V., Naugol'Naya, M. N., Verkhodanov, O. V., 1995, ARep, 39,
543.

\bibitem[Ku\-nert-\-Baj\-ra\-szew\-ska et al.(2010)]{kunert10}
Kunert-Bajraszewska, M., Gawro\'nski, M.~P., Labiano, A., Siemiginowska, A., 
2010, MNRAS, in press (Paper\,I)

\bibitem[Ku\-nert-\-Baj\-ra\-szew\-ska et al.(2006)]{kun06}
Kunert-Bajraszewska, M., Marecki, A., \& Thomasson, P. 2006, A\&A 450, 945

\bibitem[Labiano (2008a)]{labianolet} Labiano, A., 2008a, A\&A, 488, 59

\bibitem[Labiano et al.(2008b)]{labiano08} Labiano, A., O'Dea, C. P.,
Barthel, P. D., et al. 2008b, A\&A, 477, 491

\bibitem[Labiano et al.(2007)]{labiano07} Labiano, A., Barthel, P.D., O'Dea,
C.P., et al., 2007, A\&A, 463, 97

\bibitem[Labiano et al.(2005)]{labiano} Labiano, A., O'Dea, C.P., Gelderman,
R., et al., 2005, A\&A, 436, 493

\bibitem[Levesque et al.(2010)]{Levesque10} Levesque, E.~M., Kewley, L.~J., Larson, K.~L., 2010, AJ, 139, 712

\bibitem[Laing at al.(1983)]{laing83} Laing, R.~A., Riley, J.~M., \&
Longair, M.~S. 1983, MNRAS 204, 151

\bibitem[Ma\-rec\-ki\- et\- al.(2006)]{mar06} Marecki, A.,
Kunert-Bajraszewska, M., \& Spencer, R.~E., 2006, A\&A 449, 985

\bibitem[Marecki et al.(2003)]{mar03} Marecki, A., Niezgoda, J.,
W{\l}odarczak, J., Kunert, M., Spencer, R.~E., \& Kus, A.~J.,2003, PASA 20,
42

\bibitem[Massaro et al. (2010)]{Massaro10} Massaro,F., Harris, D.~E.,
Tremblay, G., Axon, D., Baum, S., Capetti, A., Chiaberge, M., Gilli, R., 
Giovannini, G., Grandi, P., Macchetto, F.~D., O'Dea, C., Risaliti, G.,
Sparks,W., 2010, arXiv:1003.2438

\bibitem[McCarthy et al. (1987)]{McCarthy87} McCarthy, P.~J.; van Breugel, W.; Spinrad, H.; Djorgovski, S. 1987, ApJ, 321, L29

\bibitem[Morganti et al.(1997)]{morganti97} Morganti, R., Tadhunter, C. N., 
Dickson, R., Shaw, M. 1997, A\&A, 326, 130

\bibitem[O'Dea et al.(2001)]{odea01} O'Dea, C.~P., et al. 2001, AJ, 121,
1915

\bibitem[O'Dea (1998)]{odea98} O'Dea, C.~P., 1998, PASP, 110, 493

\bibitem[Rawlings et al.(1989)]{raw89} Rawlings, S., Saunders, R., Eales, S. A., 
Mackay, C. D. 1989, MNRAS, 240, 701

\bibitem[Readhead et al.(1996)]{r96} Readhead, A.~C.~S., Taylor, G.~B., Xu,
W., et~al. 1996, ApJ 460, 612

\bibitem[Spergel et al.(2003)]{Spergel03} Spergel, D.~N., Verde, L., Peiris,
H.~V., Komatsu, E., Nolta, M.~R., Bennett, C.~L., Halpern, M., Hinshaw, G.,  
Jarosik, N., Kogut, A., Limon, M., Meyer, S.~S., Page, L., Tucker, G.~S., Weiland,
J.~L., Wollack, E., Wright, E.~L, 2003, ApJS, 148, 175.

\bibitem[Tasse et al.(2008)]{tasse} Tasse, C., Best, P. N., Röttgering, H.,
Le Borgne, D., 2008, A\&A, 490, 893

\bibitem[Tengstrand et al.(2009)]{Tengstrand09} Tengstrand, O., Guainazzi,
M., Siemiginowska, A., Fonseca Bonilla, N., Labiano, A., Worrall, D. M., Grandi,
P., Piconcelli, E., 2009, A\&A, 501, 89

\bibitem[Tremblay et al.(2010)]{tremblay10} Tremblay, G.~R.,
O'Dea, C.~P., Baum, S.~A., Koekemoer, A.~M., Sparks, W.~B., de Bruyn, G.,
Schoenmakers, A.~P., arXiv:1004.0388

\bibitem[de Vries et al.(1999)]{de_Vries99} de Vries, W.~H., O'Dea, C.~P., 
Baum, S.~A., Barthel, P.~D. 1999, ApJ, 526, 27.

\bibitem[de Vries et al.(1997)]{de_Vries97} de Vries, W.~H., O'Dea, C.~P., 
Baum, S.~A., Sparks, W.~B., Biretta, J., de Koff, S., Golombek, D., 
Lehnert, M. D., Macchetto, F., McCarthy, P., Miley, G. K., ApJS 110, 191.

\bibitem[de Vries et al.(2009)]{de_Vries} de Vries, N., Snellen, I. A. G.,
Schilizzi, R. T., Mack, K.-H., Kaiser, C. R., 2009, A\&A, 498, 641

\bibitem[Wilkinson et al.(1994)]{wilkinson} Wilkinson, P. N., Polatidis, A.
G., Readhead, A. C. S., Xu, W., Pearson, T. J., 1994, ApJ, 432, L87

\bibitem[Willott et al.(1999)]{willott} Willott, C. J., Rawlings, S.,
Blundell, K. M., \& Lacy, M. 1999, MNRAS, 309, 1017

\bibitem[Wright(2006)]{wright} Wright, E. L., 2006, PASP, 118, 1711.
\end{thebibliography}
\end{document}